\documentclass[12pt]{article}

\renewcommand{\sectionmark}[1]{}
\renewcommand{\subsectionmark}[1]{}
\usepackage[utf8]{inputenc}
\usepackage[top=1in, bottom=1.25in, left=1.25in, right=1in]{geometry}

\usepackage{natbib}
\usepackage{amsmath}
\usepackage{amssymb}
\usepackage{hyperref}
\usepackage{bookt abs}
\usepackage{longtable}
\usepackage[table]{xcolor}
\usepackage{bm}
\usepackage{pdflscape}
\usepackage{tabu}
\usepackage{subcaption}
\usepackage{setspace}
\usepackage{needspace}
\hypersetup{colorlinks=true,linkcolor=blue,filecolor=blue,citecolor=blue}

\usepackage{tikz}
\usepackage[labelfont=bf]{caption}
\usepackage{relsize}
\usepackage{xifthen}
\usepackage{titlesec}
\titlespacing\section{0pt}{\parskip}{\parskip}

\allowdisplaybreaks

\numberwithin{intassumption}{assumption}

\usepackage{parskip}
\expandafter\def\expandafter\normalsize\expandafter{%
    \normalsize%
    \setlength\abovedisplayskip{5pt}%
    \setlength\belowdisplayskip{5pt}%
    \setlength\abovedisplayshortskip{-8pt}%
    \setlength\belowdisplayshortskip{2pt}%
}

\title{\vspace{-3em} \fontsize{18pt}
{18pt}\selectfont The \textit{penetrance} R package for Estimation of Age Specific Risk in Family-based Studies}
\author{\fontsize{12pt}{12pt}\selectfont Nicolas Kubista$^{1,2}$, Danielle Braun$^{1,2}$, Giovanni Parmigiani$^{1,2,*}$}
\date{\fontsize{12pt}{16pt}\selectfont $^1$Department of Biostatistics, Harvard T.H. Chan School of Public Health \\
$^2$Department of Data Science, Dana Farber Cancer Institute 
\\[0.75em]
\today
\\[0.75em]
\fontsize{10pt}{10pt}\selectfont
*\textit{Email:} gp@jimmy.harvard.edu
}

\begin{document}

\maketitle

\begin{abstract}
Reliable tools and software for penetrance (age-specific risk among those who carry a genetic variant) estimation are critical to improving clinical decision making and risk assessment for hereditary syndromes.  
We introduce \textit{penetrance}, an open-source R package available on CRAN, to estimate age-specific penetrance using family-history pedigree data. The package employs a Bayesian estimation approach, allowing for the incorporation of prior knowledge through the specification of priors for the parameters of the carrier distribution. It also includes options to impute missing ages during the estimation process, addressing incomplete age information which is not uncommon in pedigree datasets. Our open-source software provides a flexible and user-friendly tool for researchers to estimate penetrance in complex family-based studies, facilitating improved genetic risk assessment in hereditary syndromes. 
\end{abstract}

\section{Introduction}

Precision prevention and early detection of heritable diseases rely on identifying individuals at increased risk. Comprehensive germline DNA analysis panels can reliably detect pathogenic germline variants (PGVs) that may increase disease risk. In this context, an accurate estimate of age-specific risk of disease is a crucial input for counseling individuals.

A penetrance function quantifies the proportion of carriers who develop a condition by a specific age. Penetrance estimation has been extensively studied in the past for a variety of study designs \citep{thomas2004statistical}. Family-based studies usually center around a proband who is typically the initial point of contact with the healthcare provider and is the individual reporting their family history. The proband is often the only genotyped individual in the family and may or may not report additional genotyping information for known relatives. Different methods for penetrance estimation have also been previously developed for family-based studies \citep{chen2009,Shin2020Penetrance}. Classical segregation analysis, for example, has been widely used to infer inheritance patterns and estimate penetrance, but it often requires large, well-characterized pedigrees and can yield imprecise estimates when sample sizes are small \citep{elston1971general, claus1991genetic}. 

However, in our assessment, there is a lack of easily usable software for penetrance estimation for family-based studies that implements a Bayesian estimation approach. For example, the widely used software package MENDEL by \cite{ Lange2001, lange2008mendel, mendel16} offers an option for penetrance estimation which can account for a full pedigree, but is difficult to use and interpret. Other penetrance estimation software and webtools are outdated and no longer maintained. To our knowledge no other software packages for penetrance estimation are readily available today. Additionally, when analyzing a study of PGV carriers, it can be important to integrate prior knowledge from published studies, which provide important information. Currently available software packages do not provide options to easily include prior knowledge. 

We introduce \textit{penetrance}, an open-source R software package, to carry out age-specific and sex-specific penetrance estimation for complex family-based studies where carrier status is not observed for all family members. The implementation is based on Bayesian estimation methods, which allows for the incorporation of prior knowledge from existing studies and provides a flexible and easy-to-use method for penetrance estimation.

Our software provides a flexible and intuitive tool for researchers to estimate penetrance in complex family-based studies, facilitating improved genetic risk assessment in hereditary syndromes. While applicable across various heritable conditions, our primary motivation stems from cancer genetics, where early detection and prevention strategies significantly impact patient outcomes \citep{Rebbeck:2018bi, Rebbeck2009, Finch2006}.

\section{Methods}
\label{sec:meth}
\subsection{Model}
Our package implements a pedigree-based Bayesian framework for estimating age-specific penetrance in family-based studies. This approach focuses on modeling disease risk for a single disease at a time, though our methods could be extended to multiple diseases in future work. 

The statistical model integrates critical population-level parameters, including carrier prevalence in the general population and baseline age-specific disease probabilities for non-carriers. 

For penetrance modeling, we employ a modified Weibull distribution with parameters $\theta = (\alpha, \beta, \gamma, \delta)$, where scale ($\alpha>0$) and shape ($\beta>0$) are standard Weibull parameters, threshold ($\delta>0$) accounts for minimum onset age, and asymptote ($0 < \gamma < 1$) represents lifetime disease probability. This parameterization creates a flexible cumulative distribution function representing disease onset probability over time. We use a parametric Bayesian approach to estimate the posterior distribution of $\theta$. Since direct calculation is intractable due to the complexity of the likelihood function and unobserved variables, we employ a fine-tuned adaptive Markov Chain Monte Carlo (MCMC) method \citep{metropolis1953equation, gilks1995markov} to approximate the posterior distribution.

\subsection{Package Usage}
\begin{figure}
    \centering
    \includegraphics[angle=+90, width=0.78\linewidth]{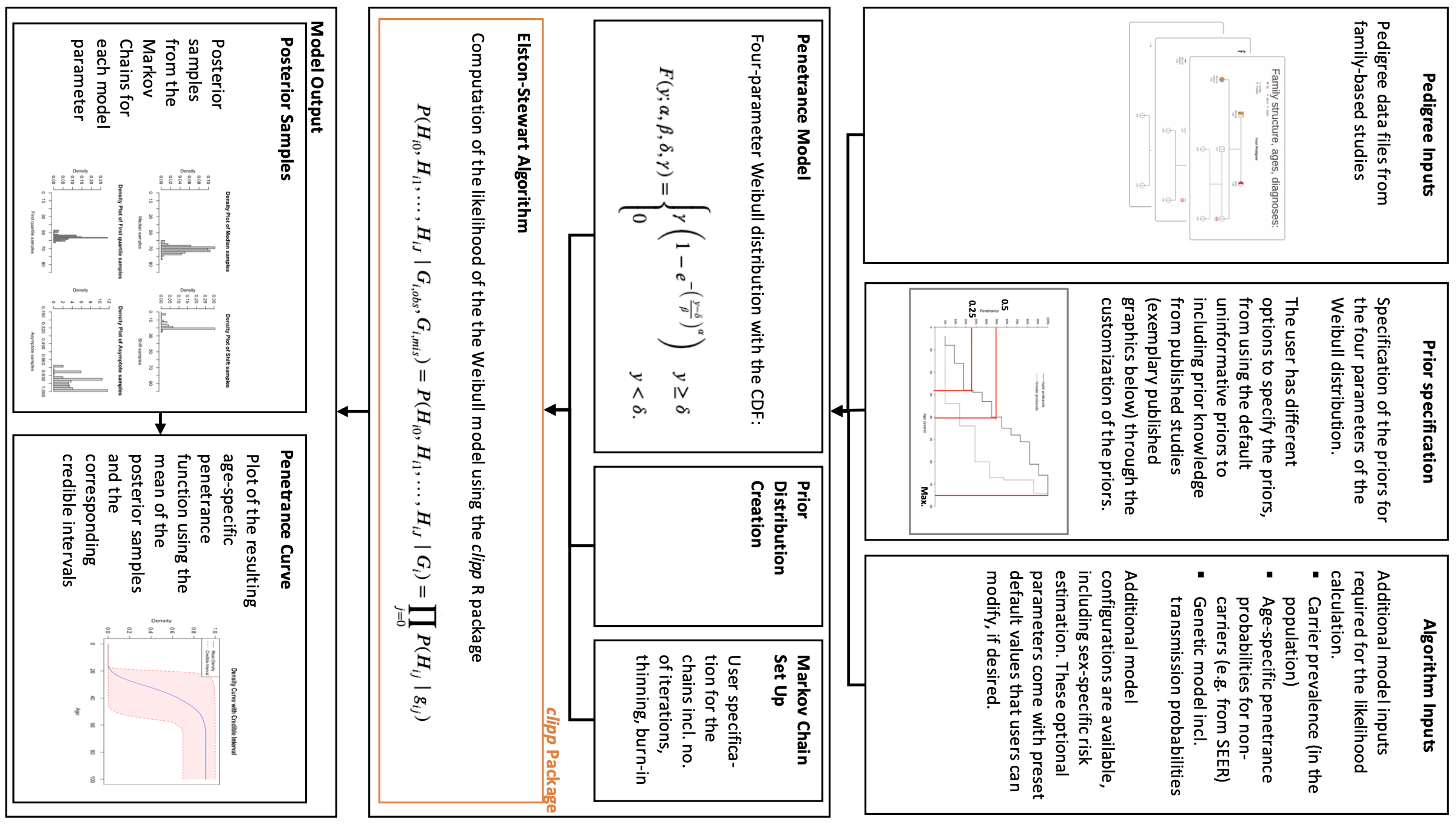}
    \caption{Workflow of the \textit{penetrance} Package with dependency on \textit{clipp}.}
    \label{fig: workflow}
\end{figure}

The methods discussed above are implemented in an easy-to-use R software package, available on CRAN. The package workflow includes three main parts: (1) the user input, including family data in the form of pedigrees, and the specifications of the options for the penetrance estimation, (2) the estimation of the posterior distribution using the MCMC approach, and (3) the output of the results in the form of the samples from the approximated posterior distribution, i.e. the estimated penetrance function. Figure \ref{fig: workflow} shows the package workflow. 

The main input is the family pedigree data reported by the proband. The pedigree file is an R data frame object. Each family tree is represented by a single data frame that includes the following columns: \textit{PedigreeID, ID, Sex, MotherID, FatherID, isProband, CurAge, isAff, Age,} and \textit{Geno}. Each row represents a unique family member, with family relationships established through parent ID references. The \textit{isProband} column (value 1) designates the individual reporting family history. Disease status is indicated through \textit{isAff} (1=diagnosed, 0=unaffected, NA=unknown) and \textit{Age} (age at diagnosis or NA). Carrier status is recorded in \textit{Geno} (1=carrier, 0=non-carrier, NA=unknown). The column \textit{CurAge} indicates the censoring age, which is the current age if the person is alive or the age at death if the person is deceased. Missing ages are allowed and can be coded as NA or left empty. More details on the structure of the pedigree required for \textit{penetrance} is available in the \textit{Fam3Pro} (formerly PanelPRO) documentation where a similar structure is used \citep{lee2021multi}.

To run the MCMC algorithm, the prior distributions for $\theta = (\alpha, \beta, \gamma, \delta)$ must be specified.  The package provides three flexible approaches to prior specification, balancing customization with ease of use. First, the package implements reasonable default uninformative priors (First Quartile: Scaled Beta(6,3), Median: Scaled Beta(2,2), Asymptote: Scaled Beta(1,1), Threshold: Uniform(5,30)), suitable for exploratory analysis when limited prior knowledge on the shape of the penetrance curves is available. Second, users can manually adjust prior parameters by modifying the
\textit{prior\_params\_default} object, which is appropriate when researchers have specific beliefs about parameter distributions that do not align with default settings. Third, the package facilitates incorporation of published penetrance estimates through several supported formats: relative risk estimates (e.g., odds ratios) that inform the asymptote parameter, detailed age-specific risk data such as information extracted from Kaplan-Meier curves, or high-level age distribution data combined with overall study size. This literature-based approach is particularly valuable when analyzing rare variants, as it leverages existing knowledge to improve estimation precision. Supplementary Materials \ref{sec: SupPriors} provides comprehensive details on the prior elicitation process and approach used to translate published data into prior distributions.

The estimation procedure follows the approach outlined in the methods section. In addition to the pedigree data ($data$) and the specified priors ($prior\_params$) the user inputs a carrier prevalence ($prev$) and a baseline age-specific probability of developing the disease ($baseline\_data$). Assuming carriers are rare, the population age-specific probability of developing the disease is a good approximation for the noncarrier penetrance. Such population-level risk data usually comes from a population registry such as the Surveillance, Epidemiology, and End Results (SEER) Program for cancer in the United States \citep{seerCRC}. 
Per default, the estimation is sex-specific, estimating two distinct set of parameters for females and males. However, the user has the option to run the estimation without considering sex, by setting \textit{sex\_specific = FALSE}.
The user can provide additional inputs to the function at run-time. An overview of the options the user can specify in the main function call can be found in the Supplementary Table \ref{tab:inputs}.

The estimation requires preparation of the pedigree data, particularly handling missing age information and adjusting the data format. A penetrance function quantifies the proportion of carriers who develop a condition by a specific age. Details on how the package handles missing ages are provided in supplementary materials.


For the estimation procedure, the initial set of parameters, denoted $\theta_0$, is drawn from the empirical distribution of the provided data, stratified by sex and disease affection status. In each iteration of the MCMC algorithm, a new proposal is generated for the parameter vector. After generating the proposal, it is essential to ensure that each parameter falls within its respective bounds. These bounds are set to maintain realistic and meaningful parameter values based on biological plausibility (see Supplementary Materials \ref{sec:SupBounds}). If any parameter falls outside these bounds, the proposal is automatically rejected, ensuring that the algorithm only considers valid parameter values for further evaluation.

Based on the proposed parameters of the Weibull distribution, \textit{penet.fn} calculates the age-specific penetrance for every individual at each iteration of the algorithm. The \textit{clipp} package is used to efficiently compute the likelihoods using the Elston-Stewart algorithm \citep{clipp}. It assumes a genetic model with a single biallelic locus. Age-specific penetrance probabilities are adjusted based on carrier status and affection status of the individual.

After running for the specified number of iterations, the \textit{penetrance} function returns a comprehensive set of outputs from the MCMC estimation process. Each chain generates its own posterior samples, and these samples collectively represent draws from the target posterior distribution. 

\section{Illustrative Example}

To illustrate our method, we conducted a simulation study designed to mirror real-world familial colorectal cancer (CRC) patterns associated with Lynch syndrome. The simulation parameters were specifically chosen to reflect characteristics of PGVs in the MLH1 gene, one of the most common mismatch repair genes associated with Lynch Syndrome. We based our carrier prevalence on population-level MLH1 data, while the penetrance parameters (for simulating disease status) were derived from established CRC risk estimates in MLH1 carriers \citep{wang2020penetrance, Felton2007}.

We simulated 130 probands with PGVs in the MLH1 gene, along with  a total  of 4,604 individuals including probands (average family size= 35.4) with 428 CRC diagnoses (203 females and 225 males). The dataset, code and resulting output of this illustration can be found in the vignette of the R package. 

We generate a chain of 20,000 posterior samples and discard the first 2,000 samples as burn-in ($burn\_in = 0.1$) in each plot. We do not perform any age imputation ($age\_imputation = FALSE$ and do not attempt any ascertainment correction by removing the proband ($remove\_proband = FALSE)$.

 The \textit{penetrance} function returns the outputs discussed in the methods section and an R object that includes the posterior samples which can be used for subsequent analysis.

\section{Conclusion}

The \textit{penetrance} open-source R package is a specialized tool for estimating age-specific disease risk in carriers of pathogenic genetic variants using family-based data. It addresses the critical need for reliable penetrance estimates in genetic counseling and risk assessment for hereditary conditions. Using a Bayesian framework with a modified Weibull distribution, the package allows researchers to incorporate prior knowledge from published studies while providing sensible defaults for those without prior information.

Key features of the package include flexible prior specification options, sex-specific penetrance estimation, and automated handling of missing age data through an optional imputation algorithm. The software implements the Elston-Stewart peeling algorithm through the clipp package for efficient likelihood calculation in complex pedigree structures. Comprehensive visualization options help researchers interpret results easily. 

The current implementation focuses on single-disease, single-carrier-status models with sex as the primary stratification variable, and assumes a unimodal Weibull distribution for age-specific penetrance patterns. While these design choices accommodate many hereditary syndromes, future extensions could incorporate competing risks, multiple disease outcomes, and more flexible penetrance distributions.

By combining statistical rigor with user-friendly implementation, \textit{penetrance} provides genetic epidemiologists with an accessible tool for estimating disease risk from family-based studies. The package was developed in the context of cancer genetics research but is applicable across many hereditary conditions where accurate risk assessment is essential for clinical decision-making. The software is freely available on CRAN with comprehensive documentation and examples.

\section{Acknowledgments}
We gratefully acknowledge support from the National Cancer Institute at the National Institutes of Health grants R01CA242218. We thank Dr. Chris Amos for providing very helpful comments. 

\section{Data Availability}
This manuscript introduces \textit{penetrance}, an innovative penetrance estimation tool. 
The package is available on CRAN and the source code can be accessed at 
\url{https://github.com/nicokubi/penetrance}.

\bibliographystyle{plainnat}
\bibliography{refs/BayesMendelBibliography, refs/GPBibliography}

\clearpage
\appendix

\setcounter{figure}{0}
\setcounter{table}{0}
\setcounter{equation}{0}

\renewcommand{\thefigure}{A\arabic{figure}}  
\renewcommand{\thetable}{A\arabic{table}}    

\begin{center}
\large{Supplementary Material for: Development of the \textit{penetrance} R package for Penetrance Estimation in Family-based Studies"
\\ by N. Kubista, D. Braun, G. Parmigiani}
\end{center}

\section{\textit{penenterance} package options}

\begin{table}[h]
\centering
\caption{\label{tab:inputs} List of model options for \textit{penetrance} (excluding prior and output options).}
\begin{tabular}{p{3cm} p{2cm} p{2cm} p{8cm}}
\toprule
Argument Name & Value & Default Value & Definition \\
\midrule
data & NA & NA & Family pedigree data frame in the appropriate format. \\
twins & List & NA & Identical twins or triplets in the family can be specified. \\
n\_chains & Integer $>$ 0 & 1 & Number of chains for parallel computation. \\
n\_iter\_per\_chain & Integer $>$ 0 & 10000 & Number of iterations per chain. \\
ncores & Integer $>$ 0 & 6 & Number of cores for parallel computation. \\
baseline\_data & Dataset & Default Object & Data for baseline risk estimates. \\
max\_age & Integer $>$ 0 & 94 & Maximum age considered for analysis. \\
remove\_proband & Boolean & FALSE & Logical indicating whether to remove probands from the analysis. \\
age\_imputation & Boolean & FALSE & Logical indicating whether to perform age imputation. \\
imp\_interval & Integer $>$ 0 & 10 & The interval at which age imputation for affected and unaffected missing ages should be performed when age\_imputation = TRUE. \\
sex\_specific & Boolean & TRUE & Logical indicating whether to perform sex-specific estimation. \\
median\_max & Boolean & TRUE & Whether to use baseline median age or max\_age as upper bound for median proposal. \\
BaselineNC & Boolean & TRUE & If TRUE, noncarrier penetrance is assumed to be the baseline penetrance. \\
var & Vector & c(0.1, 0.1, 2, 2, 5, 5, 5, 5) & Vector of initial variances for the covariance matrix of the proposal distribution. \\
burn\_in & Fraction (0 to 1) & 0 & Fraction of results to discard as burn-in. \\
thinning\_factor & Integer $\geq$ 1 & 1 & Thinning factor, applied to individual Markov chains. \\
distribution\_data & List or NA & Default Object & Data used to generate prior distributions. \\
prev & Float & 0.0001 & Prevalence (of the carrier status) in the population. \\
\bottomrule
\end{tabular}
\end{table}

\section{Prior Elicitation based on existing studies}
\label{sec: SupPriors}

\begin{figure}[htbp]
    \centering
    \begin{subfigure}[t]{0.45\linewidth}
        \centering
        \includegraphics[width=\linewidth]{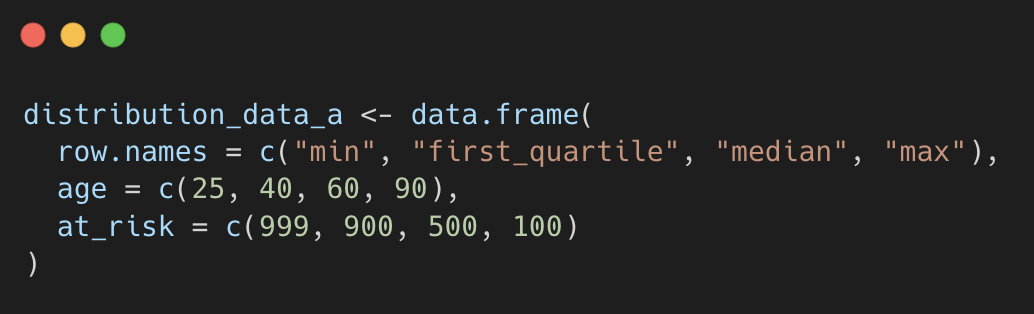}
        \caption{Example configuration with user-provided prior information on ages and resp. number of individuals at risk at those ages.}
        \label{fig:distcode-a}
    \end{subfigure}
    \hfill
    \begin{subfigure}[t]{0.45\linewidth}
        \centering
        \includegraphics[width=\linewidth]{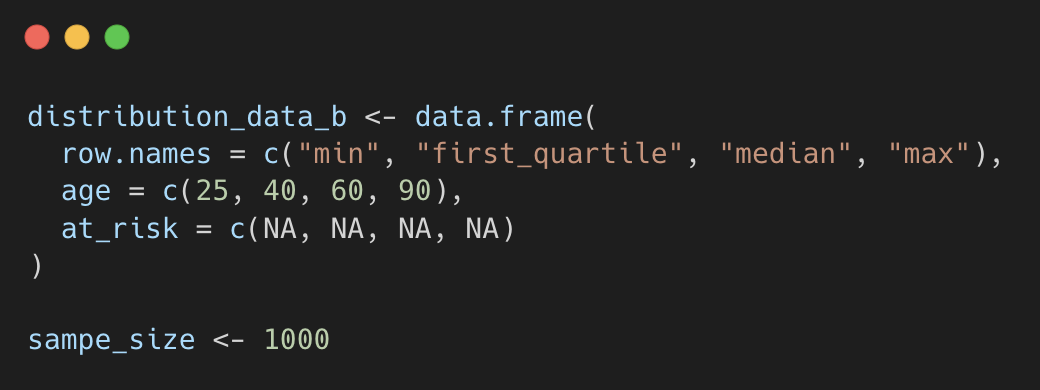}  
        \caption{Example configuration with user-provided ages and the total sample size.}
        \label{fig:distcode-b}
    \end{subfigure}
    \caption{Structure of the input object for prior information.}
    \label{fig:distcode}
\end{figure}

The package provides the user with the option to automatically generate priors based on information gathered in existing penetrance studies. Since users may often not have enough knowledge to directly define the custom parameters for the priors, we provide three options to facilitate the incorporation of different types of data that a user might be able to recover from published studies: 
 \begin{enumerate}
        \item \textbf{Relative Risk Estimates:} This option uses published measures of overall relative disease risk (e.g., odds ratios [OR] or relative risks [RR]) to inform the prior on the asymptote parameter, which represents lifetime penetrance. Specifically, if an overall disease risk measure for carriers is reported, you can enter it using the parameter $ratio$. The software then multiplies this ratio by the cumulative baseline risk to determine the mean of the asymptote parameter’s prior.
        \item \textbf{Granular Age-Specific Risk Data:} This approach requires detailed data on the age distribution of disease diagnoses in a study. Such information may be recovered, for example, from Kaplan-Meier (K-M) curves. In the $distribution\_data\_default$ object, users must populate two rows based on the information from the study: $age$ and $at\_risk$. The $age$ row should contain four time points: the earliest age of diagnosis (row name $min$), the age at which 25\% of cases occurred (row name $first\_quartile$), the age at which 50\% of cases occurred (row name $median$), and the maximum age (row name $max$). The $at\_risk$ row must contain the corresponding number of carriers still under observation at each of these ages (see Figure \ref{fig:distcode} - Panel a). Based on these inputs the algorithm then computes the parameters $\alpha$ and $\beta$ for the beta distributions of the priors for the median, first quartile, and asymptote (see Table \ref{tab: autoPriors}).
        \item \textbf{High-level Age-Specific Risk Data:} This option is suitable when only the age distribution of diagnoses and the overall study size are available (but not the number of people at risk at every time point). Users populate just the $age$ column in the $distribution\_data\_default$ object with the same four time points described above (using identical row names) and provide the total number of carriers through the $sample\_size$ parameter. The package then calculates the number of at-risk carriers at each age point using predefined proportions: 90\% of the total sample size at the first quartile age, 50\% at the median age, and 10\% at the maximum age see Figure (\ref{fig:distcode} - Panel b).
         
    \end{enumerate}

\begin{table}
\caption{\label{tab: autoPriors} Automatic prior elicitation using user inputs in distribution data default (Setting 3).}
\centering
\begin{tabular}{p{2cm} p{2cm} p{5cm} p{3cm}}
\toprule
Model Parameter & Distribution & Parameter 1 (Alpha) & Parameter 2 (Beta) \\
\midrule
First Quartile & Scaled Beta & From \textit{distribution\_data\_default}: Normalized first quartile * number of individuals at risk at the first quartile & Number of individual at risk at the first quartile - alpha \\
Median & Scaled Beta & From \textit{distribution\_data\_default}: Normalized median * number of individuals at risk at the median & Number of individual at risk at the median - alpha \\
Asymptote & Scaled Beta & From \textit{distribution\_data\_default}: Normalized max. age * number of individuals at risk at the max. age& Number of individual at risk at max. age - alpha \\ 
Threshold & Uniform &  0 & From \textit{distribution\_data\_default}: Min. age  \\ 
\bottomrule
\end{tabular}
\end{table}

\renewcommand{\thefigure}{B\arabic{figure}}  
\renewcommand{\thetable}{B\arabic{table}}    

\section{Thresholds for the Sampling of Proposals}
\label{sec:SupBounds}

Parameter bounds in our MCMC algorithm ensure biological plausibility while maintaining computational efficiency. These thresholds prevent the exploration of biologically impossible parameter combinations while ensuring interpretable results. The specific bounds are detailed in Table \ref{tab:conditions}.

\begin{table}
\centering
\begin{tabular}{p{4cm} p{10cm}}
\toprule
\textbf{Parameter} & \textbf{Condition} \\
\midrule
Asymptote & $0 \leq \text{Asymptote} \leq 1$ \\
\midrule
Threshold & $0 \leq \text{Threshold} \leq 100$ \\
\midrule
$Q_{50}$ (Median) & $Q_{50} \geq Q_{25}$ \newline
$Q_{50} \leq \text{Median Age of Onset of the SEER Baseline or}$ \newline
$Q_{50} \leq \text{max\_age}$ \\
\midrule
$Q_{25}$ (First Quartile) & $\text{Threshold} \leq Q_{25} \leq Q_{50}$ \\
\bottomrule
\end{tabular}
\caption{Conditions for the parameters irrespective of sex.}
\label{tab:conditions}
\end{table}

\section{Age Imputation}
\label{sec: ageImputation}
By default, the package assumes that the age information on probands and relatives are complete and no age imputation is required. In the case where an individual lacks an age of diagnosis or censoring age, their age-related contribution is omitted from the likelihood calculation. To address datasets with missing age data, the \textit{penetrance} package provides an option to automatically impute missing ages of diagnosis or censoring ages. The imputation is based on the individual's affection status, sex, and degree of relationship to the proband, who is typically a carrier. We address missing ages of diagnosis in affected individuals by computing posterior carrier probabilities to impute missing ages by sampling from the estimated penetrance function (for likely carriers) or the provided baseline (for likely noncarriers). For unaffected individuals with missing ages, we impute censoring ages from the empirical age distribution of unaffected individuals in the data and update these imputed ages periodically during MCMC to incorporate uncertainty into parameter estimates. The frequency with which ages are imputed affects the run time and can be specified using the option $imp\_interval$. The default is $imp\_interval = 10$, which means that age imputation is performed at every 10\textsuperscript{th} iteration. 

\section{Illustrative Example Outputs}

\begin{figure} [h!]
    \centering
    \includegraphics[width=1\linewidth]{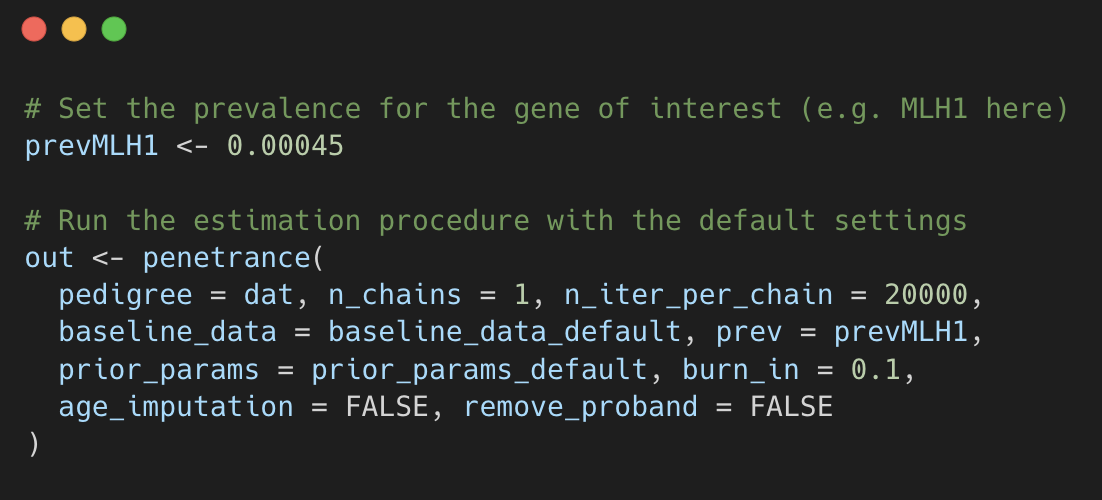}
    \caption{Exemplary code for running \textit{penetrance} for simulated data for MLH1 and CRC.}
    \label{fig:code1}
\end{figure}

\begin{figure} [h]
    \centering
    \includegraphics[width=1\linewidth]{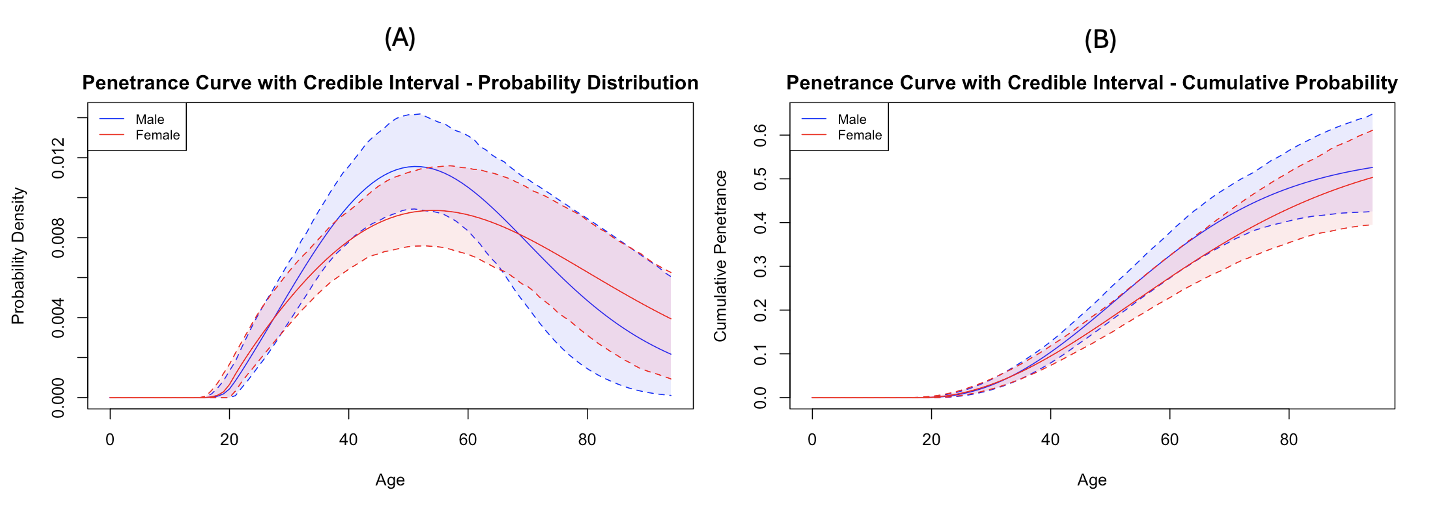}
    \caption{Age-specific absolute risk (A) and cumulative risk (B) for CRC and MLH1 for females and males with default prior parameters for simulated data.}
    \label{fig:penCurvesSIM}
\end{figure}
\end{document}